# Wide-angle emission in cylindrical moiré lattices enabled by rolling origami

Min Tang,[1,2,*] Fanzhou Lv,[2] Haiyun Dong,[3,4] Jiawei Wang,[5] Chaoyuan Jin,[6] Tun Cao,[1] Ching Hua Lee,[7] Ronny Thomale,[8] Sebastian Klembt,[9] Yana Vaynzof,[2,10] Libo Ma[2,*]

[1]School of Optoelectronics Engineering and Instrumentation Science, Dalian University of Technology, Dalian, 116024, China

[2]Leibniz Institute for Solid State and Materials Research Dresden, 01069 Dresden, Germany

[3]Key Laboratory of Photochemistry, Institute of Chemistry, Chinese Academy of Sciences, Beijing 100190, China

[4]School of Chemical Sciences, University of Chinese Academy of Sciences, Beijing 100049, China

[5]School of Integrated Circuits, Harbin Institute of Technology (Shenzhen), Shenzhen 518055, China

[6]College of Information Science and Electronic Engineering, Zhejiang University, Hangzhou 310027, China

[7]Department of Physics, National University of Singapore, 117551 Singapore, Republic of Singapore

[8]Institut für Theoretische Physik und Astrophysik, Universität Würzburg, 97074 Würzburg, Germany

[9]Technische Physik, Wilhelm-Conrad-Röntgen-Research Center for Complex Material Systems, Universität Würzburg, D-97074 Würzburg, Germany

[10]Chair for Emerging Electronic Technologies, Technical University of Dresden, 01187 Dresden, Germany

* tangminah@dlut.edu.cn, l.ma@ifw-dresden.de

ABSTRACT: Twisted photonic lattices that form moiré superlattices have attracted significant attention owing to their unique properties, such as strong field confinement and high-quality factors, in which the localized optical modes can serve as efficient light sources. However, in conventional moiré lattices, the emission direction of confined modes is typically fixed, and achieving a broad range of emission angles through simple modulation remains a significant challenge. Here, we design and fabricate single-layer moiré photonic lattices into cylindrical geometries using a nanomembrane origami technique. This approach enables wide-angle localized-mode emission while maintaining stable single-mode operation and excellent spectral uniformity. The moiré supercells support localized flat-band modes under various effective twist angles, resulting in the observation of periodic localized-mode emission over a wide range of azimuthal angles. Our research provides an approach for developing moiré light sources on curved surfaces, offering significant potential in applications that demand spatial light control, including three dimensional imaging, light detection and ranging, and topological states manipulation.

Twisted van der Waals heterostructures, formed by stacking two-dimensional atomic layers at a relative angle, have recently become an active research field due to their unique structural characteristics and the emergence of fascinating phenomena in these structures. The initial focus of related research was on graphene-based systems,[1] where twisting two layers to a specific angle leads to the occurrence of flat bands.[2, 3] These flat bands give rise to strongly correlated electronic phenomena, such as Mott insulating states and unconventional superconductivity.[4-6] This powerful "magic-angle" concept has been successfully transferred to optics by twisting two photonic crystal slabs, which form a moiré superlattice.[7-10] The resulting moiré superlattice creates a periodic potential for light, thereby introducing a new degree of freedom for manipulating electromagnetic fields. Analogous to their electronic counterparts, these twisted photonic structures provide a robust theoretical framework for investigating phenomena like the formation of photonic flat bands,[11-13] localization of photonic states,[14-17] polarization control,[18, 19] frequency tuning of the band structure,[20] and the engineering of bound states in the continuum.[21, 22]

As a key factor in photonic moiré superlattices, optical coupling between photonic crystal modes is typically achieved through the fine-tuning of the gap between photonic crystal layers[7, 23] or by employing spatial light interference.[24] A more convenient approach proposes a single layer designed as the effective stacking of two photonic crystal layers.[8] This method is favored for its simplicity and compatibility with standard planar semiconductor processes. In this type of moiré superlattice, the unique properties yielded from moiré flat bands, such as high quality ($Q$) factors and small mode volume, have led to significant applications, including the demonstration of moiré lasers[8, 25] and novel platforms for cavity quantum electrodynamics.[26,27] Although moiré photonic crystals have been extensively investigated in both fundamental research and experimental demonstrations, their spatial emission directionality remains limited due to the lack of effective strategies for tuning the fixed architecture. Although a pioneering method can tune emission direction by modulating the phase in individual unit cells,[28] however, achieving a wide range of emission angles with a simple modulation scheme remains a significant challenge.

Nanomembrane origami offers a convenient approach for creating 3D microstructures from 2D nanomembranes, providing exceptional flexibility in designing of functional micro- and nano- photonic devices.[29] Among the various microstructures created using the origami technique, self-rolled-up microtubular structures represent a versatile platform that can support efficient 3D light confinement and resonances, attracting broad scientific attention.[30] This type of microtubular photonic structure demonstrates significant potential in both fundamental physics and applications across various fields, including many-body coupling systems,[31-33] geometric phase generation,[34] in-plane to out-of-plane mode coupling,[35] optoplasmonics,[36] and laser technologies.[37] As a 3D architectural system, the microtubular structure not only offers an excellent foundation for manipulating optical fields and realizing spatially controllable light emission at the microscale, but also provides a promising route toward the development of multifunctional integrated photonic devices.

In the present work, single-layer moiré photonic lattices are designed and fabricated into a cylindrical structure using a nanomembrane origami method. This approach enables the realization of a wide-angle range of localized-mode emission due to the formation of photonic flat bands in the cylindrically structured moiré lattices. The moiré supercells can support localized flat band modes under various effective twist angles. Periodic localized-mode emission has been observed that directly corresponds to the underlying lattice periodicity. The characterization of the two-dimensional emission intensity distribution shows excellent agreement with the simulation results. A series of cylindrical moiré photonic lattices with varying effective twist angles was fabricated to investigate the wide-angle range of localized-mode emissions. The spectral consistency and intensity uniformity of the localized modes were experimentally demonstrated across different azimuthal directions in cylindrical moiré superlattices. Furthermore, the proposed system's design is compatible with batch fabrication, allowing the facet orientation of each unit cell to be flexibly designed and tuned on a chip. These cylindrical moiré superlattices thus hold great promise for enabling full angular coverage of light emission, thereby facilitating advancements in on-chip photonic applications.

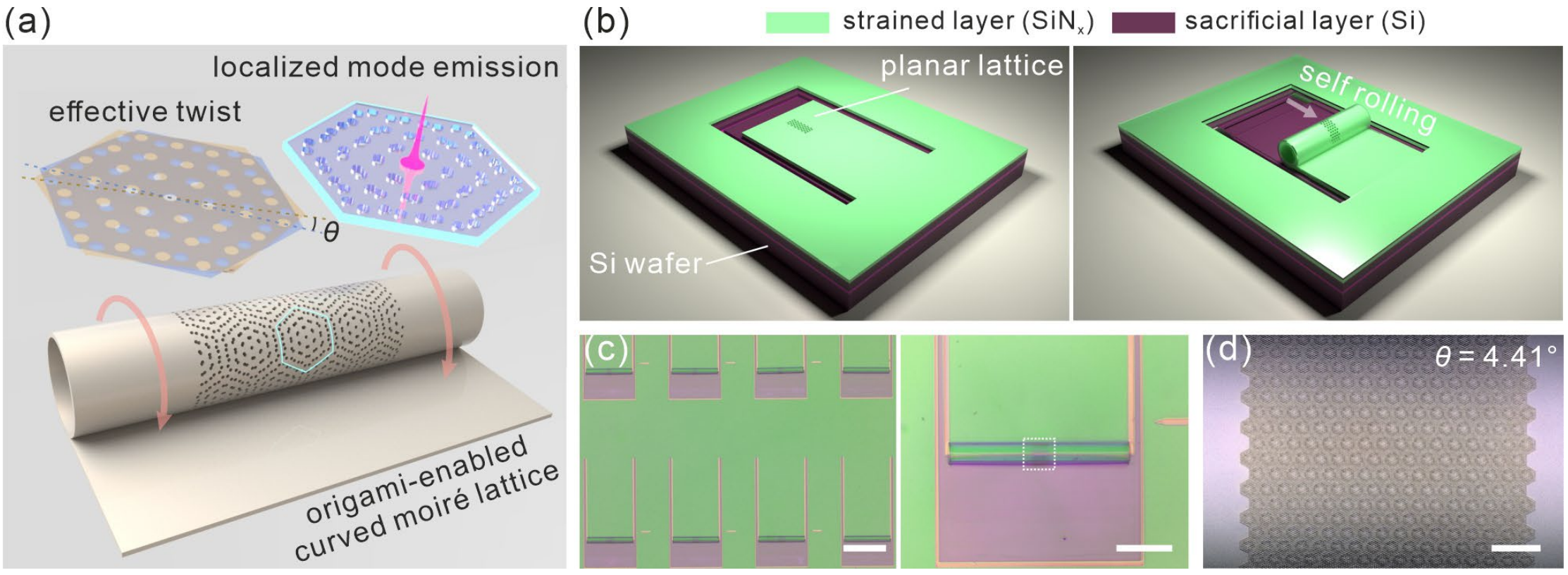

**Figure 1.** (a) Schematic diagram of moiré superlattice self-rolling into a cylindrical structure. The photonic moiré pattern is formed by effectively stacking two honeycomb photonic crystals at a twist angle $\theta$. (b) Schematic diagram showing the two primary fabrication steps of rolling up a moiré superlattice: the initial patterning of nanomembranes on a planar wafer (left panel) and the subsequent dry release of the strained nanomembrane in plasma (right panel). (c) Bright-field microscopy images of the rolled-up cylindrical structures. The region within the black frame denotes the area of the moiré lattice. (d) Optical image of a cylindrical moiré superlattice with an effective twist angle of 4.41°. The scale bars are 300, 100, and 10 μm, respectively.

As shown in Figure 1(a), the moiré pattern on a cylindrical structure is defined by two sets of triangular lattices of nanoholes. The translational periodicity of the resultant moiré pattern is determined by the angular misalignment (i.e., the effective twist angle $\theta$) between the two triangular lattices. The cylindrical moiré superlattice is fabricated by the self-rolling of a 2D moiré superlattice layer, which is triggered by releasing a pre-strained nanomembrane. As shown in Figure 1(b), the sample fabrication process consists of two primary steps: the initial patterning of nanomembranes on a planar wafer and the subsequent dry- release of the strained nanomembrane in plasma. In brief, a sacrificial layer of silicon (570 nm thick), a strained nanomembrane of $SiN_x$ (140 nm thick), and a protective layer of $Al_2O_3$ (2 nm thick) were sequentially deposited on a $SiO_2$/Si substrate. The boundaries of the rolling up of nanomembranes and the moiré lattice regions were defined using maskless lithography alignment and electron beam lithography, respectively. Subsequently, additional $Al_2O_3$ deposition and etching were performed to create an etching window in the silicon sacrificial layer. After etching away the sacrificial layer, the moiré lattice rolls up into a cylindrical structure by releasing the strain in the $SiN_x$ nanomembrane. A detailed description of the fabrication process is provided in the Supplementary Information. Figure 2(c) presents optical microscope images of the rolled-up moiré superlattices. The gray region at the center of the microtube corresponds to the area of the photonic moiré lattice. Figure 2(d) shows a moiré superlattice array with a distinct periodic pattern on the rolled-up microstructure. This self-rolling method enables the transformation of a 2D planar lattice into a curved lattice in 3D space.

To form a periodic Moiré superlattice, the twist angle $\theta$ is given by:[38, 39]

$$\theta = j\ln\left(\frac{2n_1+n_2+jn_2\sqrt{3}}{2m_1+m_2+jm_2\sqrt{3}}\right) \qquad (1)$$

In this study, a series of twist angles 4.41°, 6.01°, 5.09°, and 7.34° are considered, corresponding to $n_1 = m_2 =8, 7, 6, 5$ and $n_2 = m_1=7, 6, 5, 4$, respectively. An inverse relationship exists between the twist angle and the unit cell size, wherein larger angles yield smaller periodic structures. Differences in the unit cell size cause the localized optical modes to exhibit varying sensitivities to boundary conditions. Another critical parameter alongside the twist angle is the interlayer coupling strength. In the case of a merged single-layer moiré superlattice, this coupling strength reaches its maximum. In the context of flat band formation, interlayer coupling causes a repulsion of degenerate Bloch modes, pushing one set of the modes toward the Dirac cone and consequently forming a flat band.[8] In the single-layer moiré superlattice, key parameters concern only the twist angle and the refractive index. In the previously reported moiré lattices[8,25], the refractive indices are greater than 2.5. In this work, the $SiN_x$ membrane possesses a refractive index of around 2.0. The refractive index of the photonic crystal directly correlates with the intralayer coupling strength, with a lower refractive index resulting in a more pronounced intralayer coupling. Consequently, the relative interlayer coupling weakens. For the other key parameter, a relatively small twist angle would introduce more effective coupling sites within a single unit cell. Thus, a smaller twist angle corresponds to a flatter energy band, while a larger twist angle leads to a flat band with a broader frequency range for the localized modes. It is noteworthy that localized modes can be consistently formed with a similar distribution when varying the twist angle.

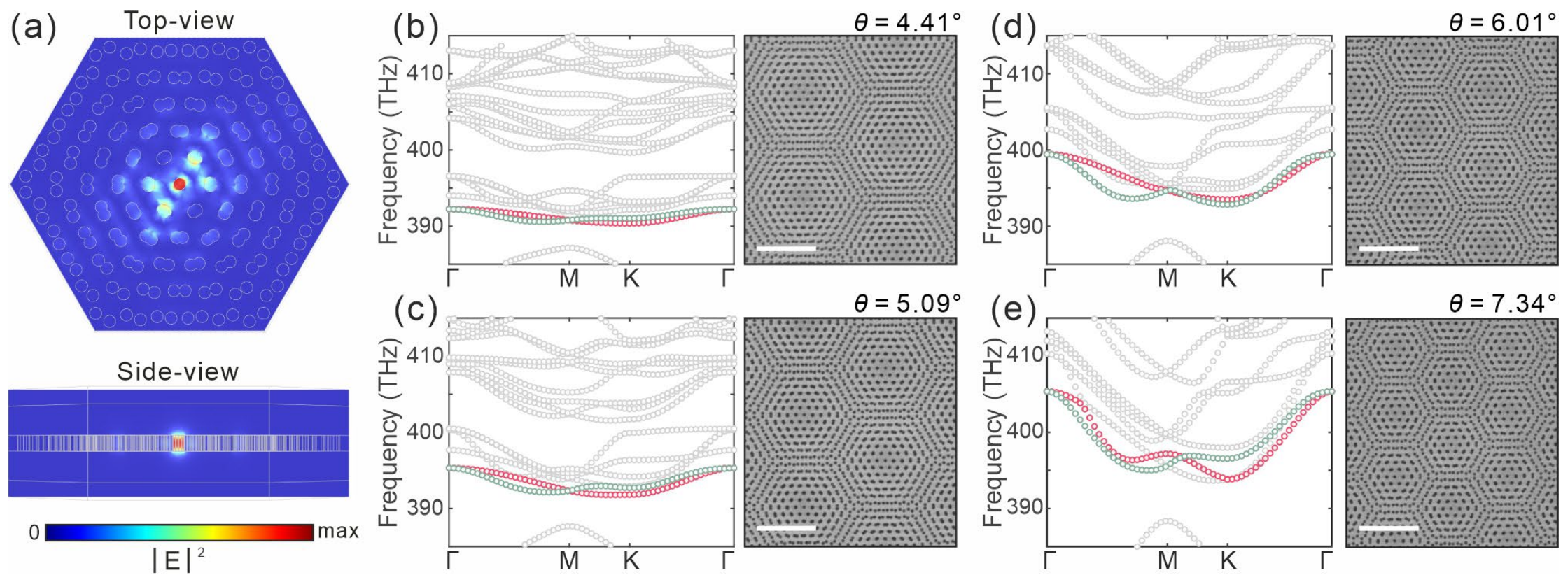

**Figure 2.** (a) 3D distribution of the localized flat band mode $|E|^2$ in a single-layer moiré superlattice at an effective twist angle of 6.01°, with cross-sectional views in the top-view and side-view planes. SEM images of the moiré superlattice at the twisted angle of (b) 4.41°, (c) 5.09°, (d) 6.01°, and (e) 7.34° and the corresponding band structures. Scale bar: 2 μm.

The COMSOL Multiphysics is utilized to simulate mode profiles and band diagrams of the moiré superlattices. The moiré microcavity with refractive index distribution $n(x,y)$ can be described by the scalar wave equation:

$$-\nabla^2\varphi = n^2\left(\frac{\omega^2}{c^2}\right)\varphi, \qquad (2)$$

where $\omega$ is the angular frequency, $c$ is the light speed in vacuum, and $\varphi$ represents the field distribution. Floquet boundary conditions are applied at the edges of the moiré superlattice. The original circular air hole possesses a radius of 55 nm and a lattice constant of 280 nm. The twist angles are set to 4.41°, 5.09°, 6.01°, and 7.34°, respectively. Three-dimensional simulations were performed to investigate the localized modes supported by the moiré superlattice. As shown in Figure 2(a), for a twist angle of 6.01°, the moiré superlattice exhibits strong optical confinement in both the in-plane and out-of-plane directions. Figures 2(b)-(e) present the SEM images of the moiré superlattices and the corresponding band structures at different twist angles. The flatness of the bands can be evaluated using the frequency $\Delta f = \omega/2\pi$. The obtained $\Delta f$ values are 1.86, 3.52, 6.61, and 11.56 THz for twist angles of 4.41°, 5.09°, 6.01°, and 7.34°, respectively, which are consistent with the expected correlation between the twist angle and flat-band formation.

For moiré superlattices with different twist angles, the mode field is primarily localized in the central air hole, exhibiting a rapid decay toward the lattice boundary. Due to the similar mode distributions, the corresponding spatial volumes or areas they occupy are nearly identical. In 2D simulations, the mode area $S_m$ is used to characterize mode confinement ability, defined as

$$S_m = \frac{\int n^2|\varphi|^2 dS}{\max(n^2|\varphi|^2)}. \qquad (3)$$

The mode areas for the flat band modes are calculated to be 1.42, 1.30, 1.17, and 1.05 × $(\lambda/n)^2$ for moiré superlattices with twist angles of 4.41°, 6.01°, 5.09°, and 7.34°, respectively, where $\lambda$ is the wavelength of the mode, and $n$ is the refractive index of $SiN_x$. This indicates that the mode can be confined to a wavelength scale in the $xy$-plane, promoting stronger light-matter interactions.

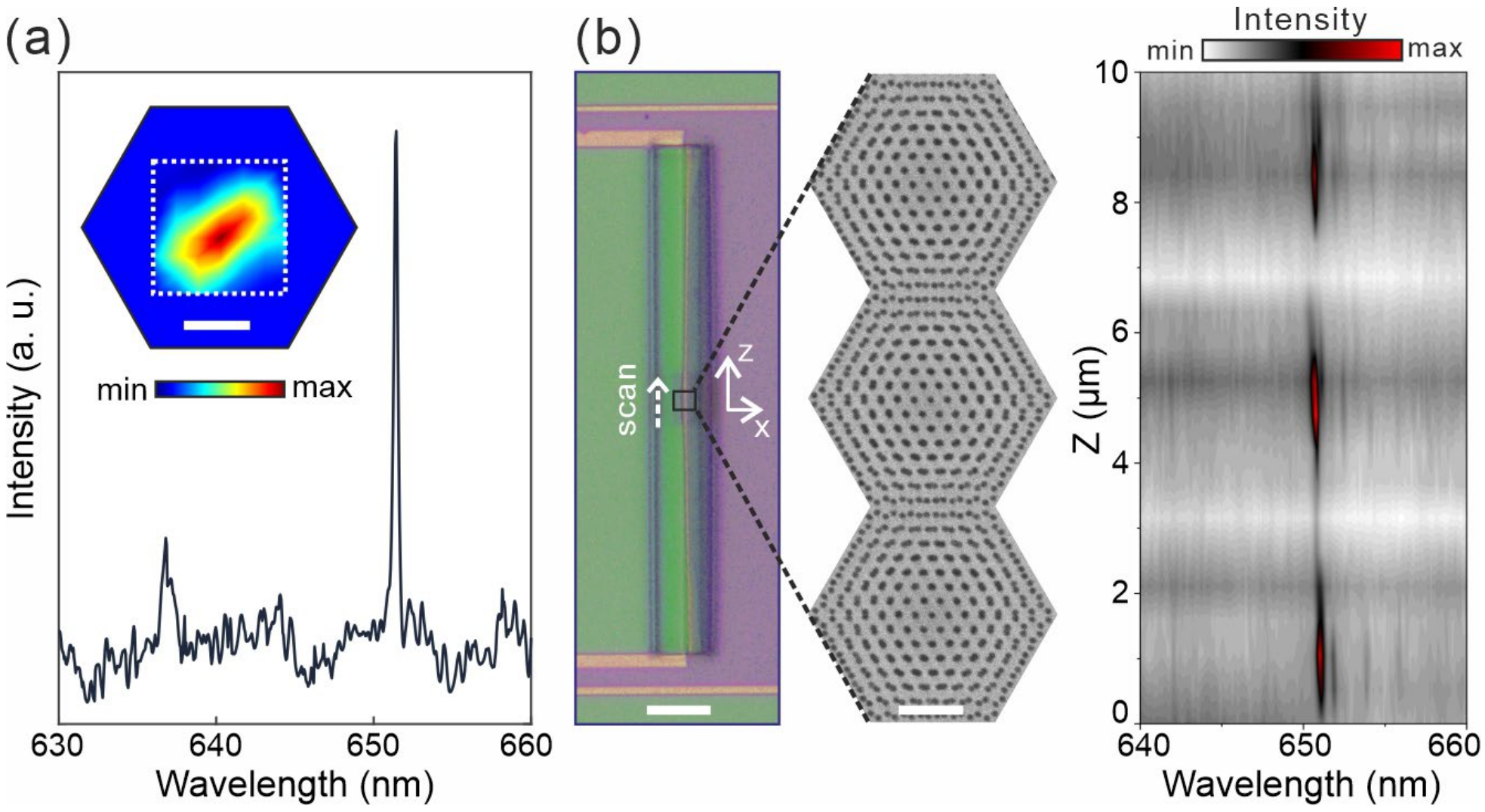

**Figure 3.** (a) Spectrum of the localized mode in a cylindrical moiré superlattice with a twist angle of 4.41°. Inset: experimental mapping of the mode within a single unit cell. The scale bar is 1 μm. (b) Spectra map of the localized mode showing its distribution across three periods of the superlattice. The scale bars from left to right are 40 μm, and 1 μm, respectively.

The cylindrical moiré lattice with a twist angle of 4.41° was systematically measured due to its well-formed flat bands and localized modes. The measurements were performed using a laser confocal system. Figure 3(a) shows a typical spectrum of the localized mode, characterized by a single prominent peak with a full width at half maximum (FWHM) of 0.28 nm, corresponding to a $Q$ factor of $2.3\times10^3$. The inset shows a two-dimensional mode intensity scan along the $x$ and $y$ directions with a mapping step of 300 nm, revealing that the mode exhibits excellent localization within an individual unit cell. To further characterize the periodicity of the mode, a line scan was conducted along the axial direction of the cylindrical structure. As illustrated in Figure 3(b), a periodic intensity distribution is revealed, with the observed period aligning precisely with the lattice periodicity.

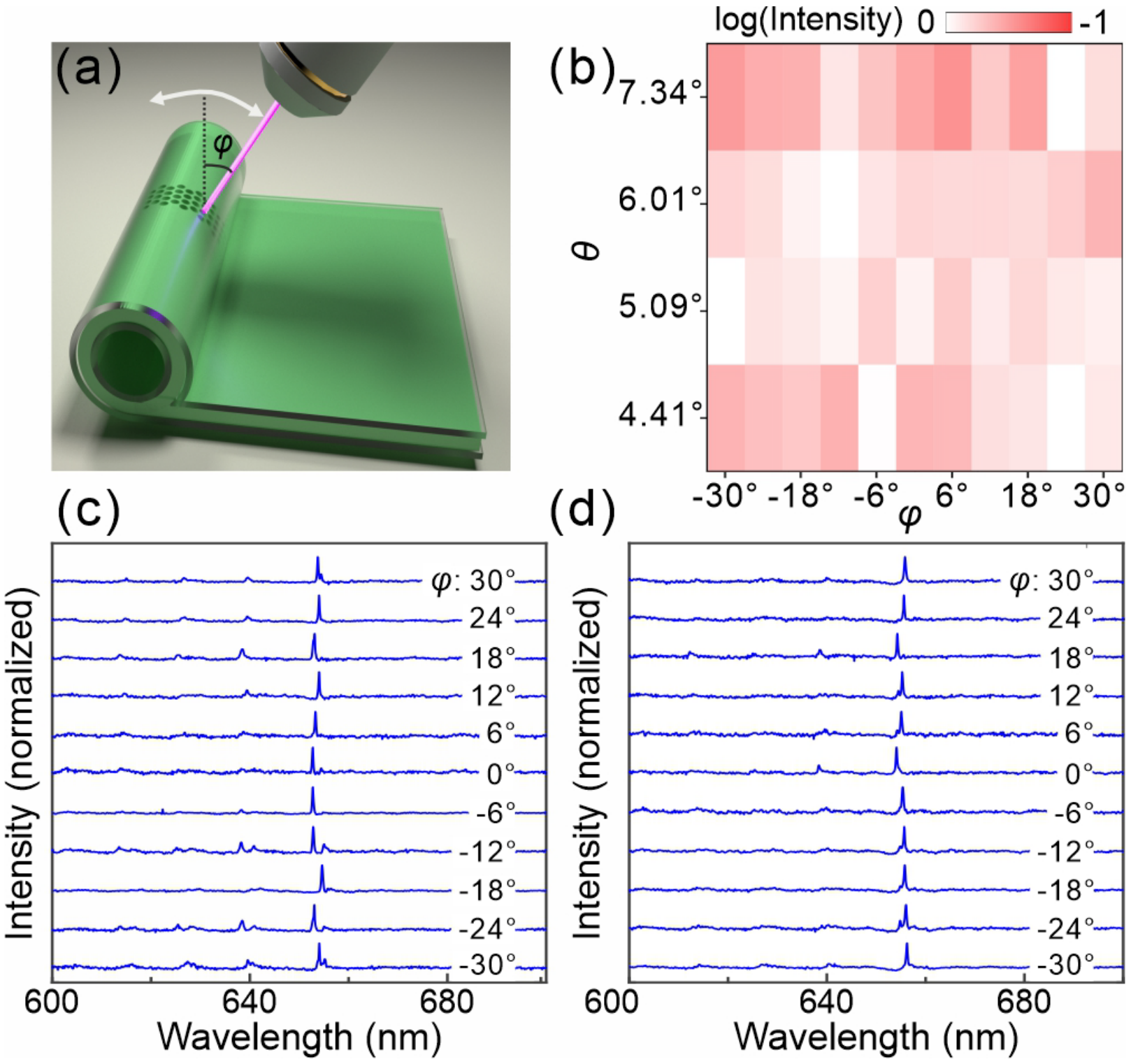

**Figure 4.** (a) Schematic illustration of angle-dependent measurement on a cylindrically structured moiré lattice. The tilt angle $\varphi$ is defined as the angle between the incident laser beam and the surface normal of the substrate. (b) Peak intensity map of localized modes measured from cylindrical moiré lattices with different twist angles $\theta$ as a function of tilt angle $\varphi$. The measured emission spectra for cylindrical moiré lattices with twist angles of (c) 4.41° and (d) 5.09° are displayed with tilt angle $\varphi$ varying from -30° to 30°. Weak sidebands are also observed in the short-wavelength range.

Angle-dependent measurements were performed on the cylindrical moiré lattices to characterize their emission behavior, as illustrated in Figure 4(a). The tilt angle $\varphi$, defined as the angle between the incident laser beam and the surface normal of the substrate, varied from -30° to 30°. Four rolled-up moiré lattices with twist angles of 4.41°, 5.09°, 6.01°, and 7.34° were investigated under these conditions. Figure 4(b) presents the peak intensity map for moiré lattices with these twist angles as a function of tilt angle $\varphi$, where the peak intensities were normalized for each sample. The results reveal distinct localized emission modes characterized by sharp peaks in the wavelength range of 600-700 nm. The FWHM measurements indicate $Q$ factors on the order of $2\times10^3$. Detailed emission spectra for each lattice structure are shown in Figure 4(c) and 4(d), and the spectra for the structures with twist angles of 6.01° and 7.34° are provided in the Supplementary Information (Figure S1(a) and (b)). In addition to the central peak, several weak sidebands are observed, originating from weakly localized higher-order dipole modes. Overall, the spectra exhibit excellent uniformity in intensity, with a variation of less than 3 dB between the strongest and weakest peaks. The Moiré lattice exhibits a reduced mode volume and exceptional single-mode characteristics owing to the low refractive index contrast and the curvature-induced suppression of nonlocal Floquet waves. Furthermore, the incorporation of gain media (such as rhodamine dyes or quantum dots) is expected to offer a promising pathway for the future development of spatial laser sources.

In summary, we have proposed and demonstrated a wide-angle range emission from localized modes in a single cylindrical moiré lattice. The system exhibits a robust localized mode with excellent spectral uniformity, as confirmed by band-structure calculations, mode profile simulations, and spatial scan measurements. The observed consistency of emission characteristics across different orientations validates the wide-angle range of localized mode emissions. Beyond establishing a versatile platform for exploring moiré photonics on curved surfaces, this work provides a foundation for the development of next-generation photonic devices with controllable spatial emission, offering potential applications in integrated optical communication, imaging, and quantum information processing.

REFERENCES

(1) Novoselov, K. S.; Mishchenko, A.; Carvalho, A.; Castro Neto, A. *Science* **2016,** 353, (6298), aac9439.

(2) Cao, Y.; Fatemi, V.; Demir, A.; Fang, S.; Tomarken, S. L.; Luo, J. Y.; Sanchez-Yamagishi, J. D.; Watanabe, K.; Taniguchi, T.; Kaxiras, E.; Ashoori, R. C.; Jarillo-Herrero, P. *Nature* **2018,** 556, (7699), 80-84.

(3) Cao, Y.; Fatemi, V.; Fang, S.; Watanabe, K.; Taniguchi, T.; Kaxiras, E.; Jarillo-Herrero, P. *Nature* **2018,** 556, (7699), 43-50.

(4) Chen, G.; Sharpe, A. L.; Gallagher, P.; Rosen, I. T.; Fox, E. J.; Jiang, L.; Lyu, B.; Li, H.; Watanabe, K.; Taniguchi, T.; Jung, J.; Shi, Z.; Goldhaber-Gordon, D.; Zhang, Y.; Wang, F. *Nature* **2019,** 572, (7768), 215-219.

(5) Uri, A.; de la Barrera, S. C.; Randeria, M. T.; Rodan-Legrain, D.; Devakul, T.; Crowley, P. J. D.; Paul, N.; Watanabe, K.; Taniguchi, T.; Lifshitz, R.; Fu, L.; Ashoori, R. C.; Jarillo-Herrero, P. *Nature* **2023,** 620, (7975), 762-767.

(6) Tang, H.; Wang, Y.; Ni, X.; Watanabe, K.; Taniguchi, T.; Jarillo-Herrero, P.; Fan, S.; Mazur, E.; Yacoby, A.; Cao, Y. *Nature* **2024,** 632, (8027), 1038-1044.

(7) Tang, H.; Lou, B.; Du, F.; Zhang, M.; Ni, X.; Xu, W.; Jin, R.; Fan, S.; Mazur, E. *Science Advances* **2023,** 9, (28), eadh8498.

(8) Mao, X. R.; Shao, Z. K.; Luan, H. Y.; Wang, S. L.; Ma, R. M. *Nat Nanotechnol* **2021,** 16, (10), 1099-1105.

(9) Wang, P.; Zheng, Y.; Chen, X.; Huang, C.; Kartashov, Y. V.; Torner, L.; Konotop, V. V.; Ye, F. *Nature* **2020,** 577, (7788), 42-46.

(10) Meng, Y.; Feng, J.; Han, S.; Xu, Z.; Mao, W.; Zhang, T.; Kim, J. S.; Roh, I.; Zhao, Y.; Kim, D.-H.; Yang, Y.; Lee, J.-W.; Yang, L.; Qiu, C.-W.; Bae, S.-H. *Nature Reviews Materials* **2023,** 8, (8), 498-517.

(11) Dong, K.; Zhang, T.; Li, J.; Wang, Q.; Yang, F.; Rho, Y.; Wang, D.; Grigoropoulos, C. P.; Wu, J.; Yao, J. *Phys Rev Lett* **2021,** 126, (22), 223601.

(12) Tang, H.; Du, F.; Carr, S.; DeVault, C.; Mello, O.; Mazur, E. *Light Sci Appl* **2021,** 10, (1), 157.

(13) Tang, H.; Ni, X.; Du, F.; Srikrishna, V.; Mazur, E. *Applied Physics Letters* **2022,** 121, (23).

(14) Fu, Q.; Wang, P.; Huang, C.; Kartashov, Y. V.; Torner, L.; Konotop, V. V.; Ye, F. *Nature Photonics* **2020,** 14, (11), 663-668.

(15) Nguyen, D. X.; Letartre, X.; Drouard, E.; Viktorovitch, P.; Nguyen, H. C.; Nguyen, H. S. *Physical Review Research* **2022,** 4, (3).

(16) Ni, X.; Liu, Y.; Lou, B.; Zhang, M.; Hu, E. L.; Fan, S.; Mazur, E.; Tang, H. *Phys Rev Lett* **2024,** 132, (7), 073804.

(17) Saadi, C.; Nguyen, H. S.; Cueff, S.; Ferrier, L.; Letartre, X.; Callard, S. *Optica* **2024,** 11, (2).

(18) Gromyko, D.; An, S.; Gorelik, S.; Xu, J.; Lim, L. J.; Lee, H. Y. L.; Tjiptoharsono, F.; Tan, Z. K.; Qiu, C. W.; Dong, Z.; Wu, L. *Nat Commun* **2024,** 15, (1), 9804.

(19) Lou, B.; Fan, S. *ACS Photonics* **2022,** 9, (3), 800-805.

(20) Lou, B.; Wang, B.; Rodríguez, J. A.; Cappelli, M.; Fan, S. *Science Advances* **2022,** 8, (48), eadd4339.

(21) Qin, H.; Chen, S.; Zhang, W.; Zhang, H.; Pan, R.; Li, J.; Shi, L.; Zi, J.; Zhang, X. *Nat Commun* **2024,** 15, (1), 9080.

(22) Zhang, T.; Dong, K.; Li, J.; Meng, F.; Li, J.; Munagavalasa, S.; Grigoropoulos, C. P.; Wu, J.; Yao, J. *Nat Commun* **2023,** 14, (1), 6014.

(23) Tang, H.; Lou, B.; Du, F.; Gao, G.; Zhang, M.; Ni, X.; Hu, E.; Yacoby, A.; Cao, Y.; Fan, S. *Nature Photonics* **2025**, 1-8.

(24) Wang, P.; Fu, Q.; Konotop, V. V.; Kartashov, Y. V.; Ye, F. *Nature Photonics* **2024,** 18, (3), 224-229.

(25) Raun, A.; Tang, H.; Ni, X.; Mazur, E.; Hu, E. L. *ACS Photonics* **2023,** 10, (9), 3001-3007.

(26) Yan, S.; Li, H.; Yang, J.; Chen, X.; Liu, H.; Dai, D.; Zhu, R.; Ma, Z.; Shi, S.; Yang, L. *Nature Communications* **2025,** 16, (1), 1-8.

(27) Wang, Y.-T.; Ye, Q.-H.; Yan, J.-Y.; Qiao, Y.; Liu, Y.-X.; Ye, Y.-Z.; Chen, C.; Cheng, X.-T.; Li, C.-H.; Zhang, Z.-J. *Science Advances* **2025,** 11, (21), eadv8115.

(28) Luan, H. Y.; Ouyang, Y. H.; Zhao, Z. W.; Mao, W. Z.; Ma, R. M. *Nature* **2023,** 624, (7991), 282-288.

(29) Smela, E.; Inganäs, O.; Lundström, I. *Science* **1995,** 268, (5218), 1735-1738.

(30) Schmidt, O. G.; Eberl, K. *Nature* **2001,** 410, (6825), 168-168.

(31) Wang, X.; Wang, Z.; Dong, H.; Saggau, C. N.; Tang, H.; Tang, M.; Liu, L.; Baunack, S.; Bai, L.; Liu, J.; Yin, Y.; Ma, L.; Schmidt, O. G. *Nano Lett* **2022,** 22, (16), 6692-6699.

(32) Saggau, C. N.; Valligatla, S.; Wang, X.; Dong, H.; Ma, L.; Schmidt, O. G. *Laser & Photonics Reviews* **2022,** 16, (9).

(33) Wang, J.; Yin, Y.; Hao, Q.; Zhang, Y.; Ma, L.; Schmidt, O. G. *Advanced Optical Materials* **2017,** 6, (1).

(34) Ma, L. B.; Li, S. L.; Fomin, V. M.; Hentschel, M.; Gotte, J. B.; Yin, Y.; Jorgensen, M. R.; Schmidt, O. G. *Nat Commun* **2016,** 7, 10983.

(35) Valligatla, S.; Wang, J.; Madani, A.; Naz, E. S. G.; Hao, Q.; Saggau, C. N.; Yin, Y.; Ma, L.; Schmidt, O. G. *Advanced Optical Materials* **2020,** 8, (22).

(36) Yin, Y.; Li, S.; Böttner, S.; Yuan, F.; Giudicatti, S.; Saei Ghareh Naz, E.; Ma, L.; Schmidt, O. G. *Physical Review Letters* **2016,** 116, (25), 253904.

(37) Dong, H.; Saggau, C. N.; Zhu, M.; Liang, J.; Duan, S.; Wang, X.; Tang, H.; Yin, Y.; Wang, X.; Wang, J.; Zhang, C.; Zhao, Y. S.; Ma, L.; Schmidt, O. G. *Advanced Functional Materials* **2021,** 31, (51).

(38) Oudich, M.; Kong, X.; Zhang, T.; Qiu, C.; Jing, Y. *Nat Mater* **2024,** 23, (9), 1169-1178.

(39) Martí-Sabaté, M.; Torrent, D. *Physical Review Applied* **2021,** 15, (1), L011001.

**Experimental Section**

*Substrates*: 10 mm ×10 mm silicon on insulator (SOI) wafer with 2 μm buried thermal oxide (BOX) is used as the substrate.

*Deposition*: The deposition membrane structure consists of an $Al_2O_3$ stop layer/Si sacrificial layer/$Al_2O_3$ protecing layer/$SiN_x$ strained layer from bottom to top. Both $Al_2O_3$ layers are 4 nm and were deposited by ALD (GEMStar XT, Arradiance LLC, Littleton, USA). Silicon layer functions as sacrifice layer and was deposited by plasma enhanced chemical vapor deposition (PECVD, SI 500 D, Sentech Instruments). The 140 nm $SiN_x$ strained layer was deposited by the same PECVD tool while the strain is introduced by changing of the bias and ICP-power (ICP: 1200 W, 150V, 30 s,10 s/ICP:100 W 0 V, 230 s) as process gases silane and nitrogen were used ($SiH_4$ [5% in He]: 250 sccm, $N_2$: 80/8/80 sccm, Ar: 140 sccm, Time: 30s/10s/230s). After the MLA/EBL and following RIE process, another 2 nm $Al_2O_3$ layer was deposited for protecting $SiN_x$ during sacrifice layer etching.

*Patterning*: AZ5214E (photoresist, Mircochemicals GmbH, Ulm, Germany) was spun onto the samples at 4500 rpm, which gives a thickness of 1.3 μm of the photoresist. Pattering was performed by a maskless aligner (MLA 100, Heidelberg Instruments Mikrotechnik GmbH, Heidelberg, Germany) at a dose of 90 mJ $cm^{-2}$. Then the sample was developed in AZ 726 MIF developer (Microchemicals) for 45s. SML 300 (EBL resist, EM resist LTD, Macclesfield, UK) was spun onto the samples (1700/6000 rpm, 60/3 s). The sample was exposed at 50 kV in an EBL system (Voyager, Raith, GmbH, Dortmund, Germany) utilizing a dose of 800 μC $cm^{-2}$. Then the sample was developed in AR 600-56 (Allresist) for 90 s (including 50 sultrasonic cleaning).

*Etching*: The moiré structures were etched in an ICP-RIE (Plasma Lab 100, Oxford Instruments plc, Abingdon, UK) with $CHF_3$: 20 sccm, $CF_4$:10 sccm, and $O_2$:6 sccm as reactive gases. ICP: 450 W, Bias:145 V, pressure: 0.01 mbar, 80 s. The membranes were structured by photolithography and ICP-RIE etching. For $Al_2O_3$ layer it was etched with $BCl_3$: 20 sccm, He:10 sccm, ICP: 400 W, Bias:80 V, pressure: 0.01 mbar, 60 s. For $SiN_x$ layer it was etched with $SF_6$: 20 sccm, He:10 sccm, ICP: 400 W, Bias:15 V, pressure: 0.01 mbar, 180 s.

*Release*: Samples were released by etching Si sacrificial layer inside a xenon difluoride etching system (Xactix® e2; Orbotech Ltd., Yavne, Israel).

*Sample Characterization*: Scanning Electron Microscopy images were taken on a DSM982 GEMINI at 5 kV, (Carl Zeiss Microscopy GmbH, Jena, Germany), Optical characterization was performed with a confocal photoluminescence setup (LabRAM HR Evolution, Horiba Ltd., Kyoto, Japan). A laser beam at an excitation wavelength of 457 nm was focused onto the microtube cavities surface through a long working distance objective lens.

**Emission spectra of moiré microtubes with different titled angles**

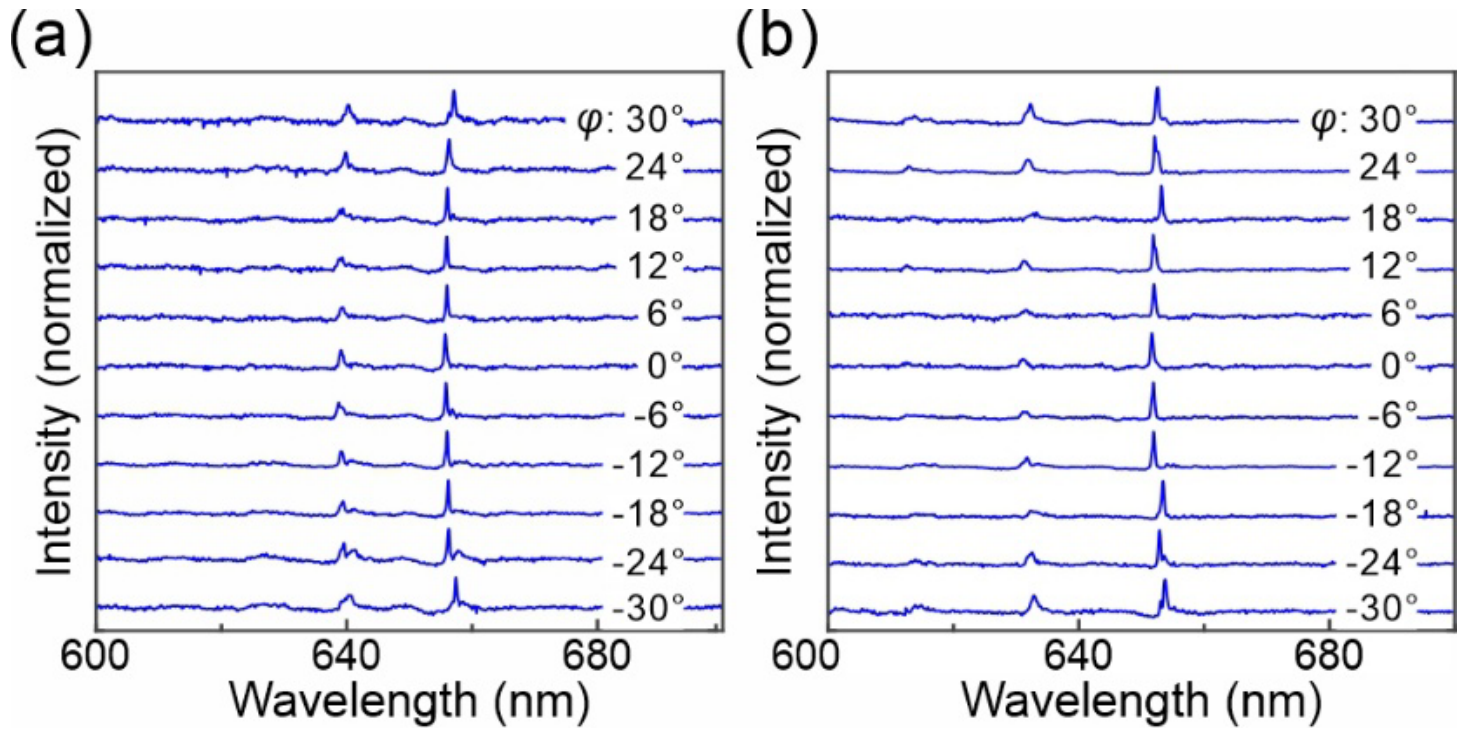

**Figure S1.** Detailed emission spectra of moiré microtubes with twist angles of (a) 6.01° and (b) 7.34°, measured with $\varphi$ varying from -30° to 30°.